\documentclass[a4paper,reqno]{amsart}
\usepackage{amssymb}
\newtheorem{theorem}{Theorem}
\newtheorem{lemma}{Lemma}

\newcommand{\bfp}{{\bf p}}

\newcommand{\bfx}{{\bf x}}

\newcommand{\bfe}{{\bf e}}
\newcommand{\bfA}{{\bf A}}
\newcommand{\bfB}{{\bf B}}
\newcommand{\bfalpha}{\mbox{\boldmath $ \alpha$}}

\newcommand{\bfsigma}{\mbox{\boldmath $ \sigma$}}

\newcommand{\bfnabla}{\mbox{\boldmath $ \nabla$}}

\title[High-field limit of the ionic  ground state] 
{The ground state of relativistic ions in the limit of high
magnetic fields}
\author{D. H. Jakuba\ss a-Amundsen  }
\address{D.H. Jakuba\ss a-Amundsen \\Mathematics Institute\\
University of Munich\\ Theresienstr. 39\\ 80333
Munich, Germany } \email{dj@mathematik.uni-muenchen.de
}

\begin{document}
\bibliographystyle{plain}
\begin{abstract}
We consider the pseudorelativistic no-pair Brown-Ravenhall operator
for the description of relativistic one-electron ions in a homogeneous magnetic field $\bfB$.
It is shown for central charge $Z\leq 87$ that their ground state energy decreases according to $\sqrt{B}$ as $B \to \infty$, 
in contrast to the nonrelativistic behaviour.
\end{abstract}
\maketitle


\vspace{0.5cm}

\section{Introduction}

A relativistic atomic electron of mass $m$ in a magnetic field $\bfB=\bfnabla \times \bfA$ resulting from a vector potential $\bfA$ is described by the Dirac operator $H$ \cite{Th},
\begin{equation}\label{1.1}
H\;=\;D_A+V,\qquad\quad D_A\;=\;\bfalpha \bfp_A\,+\beta m,
\end{equation}
where $\bfp_A=\bfp-e\bfA,\;\,\bfalpha, \beta$ are Dirac matrices and $V=-\frac{\gamma}{x}$ is the Coulomb field generated by a point nucleus of charge $Z$ fixed at
the origin.
The coordinate and momentum of the electron are denoted, respectively, by $\bfx$ and $\bfp\;$ (with $x=|\bfx|=\sqrt{x_1^2+x_2^2+x_3^2}\,)$,
and the field strength is $\gamma =Ze^2.$ Relativistic units ($\hbar =c =1$) are used, with $e^2\approx 1/137.04$ being the fine structure constant.

A widely used pseudorelativistic operator which  accounts for the spin degrees of freedom is the Brown-Ravenhall operator $h^{BR}.$
It can be obtained from a projection of $H$ onto the positive spectral subspace of the electron at $V=0$ \cite{BR} (see also \cite{EPS} for its mathematical analysis).

The Brown-Ravenhall operator in a magnetic field \cite{DeV,LSS,Jaku06} is given by
\begin{equation}\label{1.2}
h^{BR}\;=\;E_A+V_1+V_2
\end{equation}
$$V_1\;=\;-\gamma\,A_E\,\frac{1}{x}\,A_E,\qquad V_2\;=\; -\gamma \,A_E\;\frac{\bfsigma\bfp_A}{E_A+m}\;\frac{1}{x}\;\frac{\bfsigma\bfp_A}{E_A+m}\,A_E,$$
where $E_A=|D_A|$ is the kinetic energy operator,
\begin{equation}\label{1.3}
E_A\;=\;\sqrt{(\bfsigma\bfp_A)^2+m^2}\;=\;\sqrt{p_A^2-e\bfsigma\bfB+m^2}\,\geq m,\qquad A_E\;=\;\sqrt{\frac{E_A+m}{2E_A}}.
\end{equation}
Here, $\bfsigma = (\sigma_1,\sigma_2,\sigma_3) $ is the vector of Pauli spin matrices.
 $h^{BR}$ acts in the Hilbert space $L_2({\Bbb R}^3) \otimes {\Bbb C}^2$ and 
its form domain
 is $H_{1/2}({\Bbb R}^3) \otimes {\Bbb C}^2$ where $H_{1/2}$ denotes a Sobolev space.
For $\bfB$ bounded or $\bfB \in L_2({\Bbb R}^3)$ it has been shown 
\cite{Jaku06,Jaku09} that the potential $V_1+V_2$ is relatively form bounded with respect to $E_A$,
\begin{equation}\label{1.4}
|(\psi,(V_1+V_2)\;\psi)|\;\leq\; \frac{\gamma}{\gamma_c}\;(\psi, E_A\,\psi)\;+\;\gamma\;c_B\;\|\psi\|^2
\end{equation}
for $\psi \in H_{1/2}({\Bbb R}^3) \otimes {\Bbb C}^2,\;$
where $\gamma_c=\frac{2}{\pi},\;c_B\,=\frac{\pi}{2}\,\sqrt{e\|\bfB\|_\infty},$ respectively $\gamma_c=\frac{2}{\pi} -\epsilon,\;c_B=c(\epsilon)\,\|\bfB\|_2^2$ with 
$\epsilon >0$ arbitrarily small and some  constant $c(\epsilon)$ for $\bfB$ bounded respectively $\bfB \in L^2({\Bbb R}^3).$
For subcritical potential strength, $\gamma <\gamma_c,$ the form bound is smaller than one such that $h^{BR}$
is bounded from below, allowing for its extension to a self-adjoint operator.
Most recently \cite{MS} it was shown that for locally bounded $\bfA$ the 
critical potential strength for the semiboundedness of $h^{BR}$ could be increased to $ \tilde{\gamma}_{c}=2/(\frac{\pi}{2}+\frac{2}{\pi})$. 
However, no explicit $\bfB$-dependence of the lower bound of $h^{BR}$ was provided.

In the following we take $\bfB=B\bfe_3$ to be a constant magnetic field along the $\bfe_3$-axis,  generated by
\begin{equation}\label{1.5}
\bfA(\bfx)\;=\;\frac{B}{2}\;(-x_2,x_1,0)
\end{equation}
which obeys $\bfnabla \cdot \bfA=0.$

A matter of interest is the behaviour of the ground state energy
 when $B \to \infty$, which until now was only
rigorously studied for systems with small nuclear charge.
In this nonrelativistic limit the Dirac operator $H$ turns into the Pauli operator $H_P=(\bfsigma\bfp_A)^2+V.$
For this operator, using a scaling property by which
 the pair $(B,\,\frac{\gamma}{x})$ changes into $(B_0,\,\frac{\lambda}{x})$ where $B_0$ is a fixed field and $\lambda \to 0$ as $B \to \infty$,
the ground-state asymptotics could be investigated by means of a perturbative treatment of the electric potential \cite{AHS3}.
It was derived that, as conjectured earlier \cite{RS}, the ground-state energy  decreases according to $(\ln B)^2$ as $B \to \infty$, the error being of the
order of $\ln B \cdot \ln(\ln B).\;$
Recently the Dirac operator itself was studied
for finite magnetic fields and small nuclear charge (including the simultaneous limits
$B \to \infty$ and $Z \to 0$) in the context of the lowest bound state's diving into the negative continuum for sufficiently large  $B$ \cite{DEL}. A
similar scaling as in \cite{AHS3} was introduced, and an upper and lower bound for the ground-state energy
was provided which behave like $B^\frac12 $.

As concerns the Brown-Ravenhall operator $h^{BR}$ its ground state was estimated with the help
of a variational wavefunction, similar to the one used in the Schr\"{o}dinger case \cite{RMS}, with the result that the variationally determined ground-state energy
decreases asymptotically like $B^\frac12$ \cite{Jaku09}. 
Recalling that the variational energy is an upper bound while the form boundedness (\ref{1.4})  provides a lower bound,
the ground-state energy of $h^{BR}$ is thus, like that of the Dirac operator, sandwiched between bounds which decrease like $B^\frac12$ for  $B \to \infty$.

The aim of the present work is to prove a stronger statement.

\begin{theorem}\label{t.1}
Let $h^{BR}=E_A+V_1+V_2$ be the Brown-Ravenhall operator for an electron in a high  central
Coulomb field of strength $0.1 \lesssim\gamma <\frac{2}{\pi}\;\;(Z\leq 87)$ and in a homogeneous magnetic field $\bfB.$
Then the ground-state energy behaves like
\begin{equation}\label{1.6}
E_g\;\sim\;-c\;\sqrt{B}\qquad\qquad (B \to \infty)
\end{equation}
where $c>0$ is some constant.
\end{theorem}

\section{Scaling property}
\setcounter{equation}{0}

Let us apply the scaling introduced by \cite{AHS3} to the Brown-Ravenhall operator, and set $B=:\mu_0B_0$ with $\mu_0>0$ and $B_0$ some fixed constant field.
Then $\mu_0 \to \infty$ as $B \to \infty.$
Define $\tilde{x}_k:= \mu_0^{\frac12}x_k\;\;(k=1,2,3)\;$ such that
$$p_k\;=\;-i\;\partial_{x_k}\;=\;-i\;\mu_0^{\frac12}\,\partial_{\tilde{x}_k}\;=\; \mu_0^{\frac12}\,\tilde{p}_k$$
\begin{equation}\label{2.1}
\bfA(\bfx)\;=\;\frac{\mu_0B_0}{2}\;(-\mu_0^{-\frac12}\,\tilde{x}_2,\,\mu_0^{-\frac12}\,\tilde{x}_1,0)\;=: \mu_0^{\frac12}\,\bfA_0(\tilde{\bfx}).
\end{equation}

Then with $m/\sqrt{\mu_0}\,=:\tilde{m},$
\begin{equation}\label{2.2}
h^{BR}\;=\;\mu_0^{\frac12}\,(\tilde{E}_A+\tilde{V}_1+\tilde{V}_2)\;=: \mu_0^{\frac12}\,\tilde{h}^{BR}
\end{equation}
where
$$\tilde{E}_A:=\;\sqrt{\tilde{p}_A^2-e\bfsigma B_0\,+\tilde{m}^2},\qquad \tilde{\bfp}_A\;=\;\tilde{\bfp}\,-\,e\bfA_0(\tilde{\bfx})$$
\begin{equation}\label{2.3}
\tilde{V}_1:=\;-\gamma\,A_{\tilde{E}}\,\frac{1}{\tilde{x}}\,A_{\tilde{E}},\qquad
\tilde{V}_2:=\;-\gamma A_{\tilde{E}}\;\frac{\bfsigma \tilde{\bfp}_A}{E_{\tilde{A}}+\tilde{m}}\;\frac{1}{\tilde{x}}\;\frac{\bfsigma \tilde{\bfp}_A}{E_{\tilde{A}}+\tilde{m}}\;A_{\tilde{E}}
\end{equation}
and $A_{\tilde{E}}$ follows from (\ref{1.3}) where $E_A,m$ is replaced by $\tilde{E}_A,\tilde{m}.$
If we introduce the scaled ground-state energy $\tilde{E}_g:= \mu_0^{-\frac12}\,E_g$ then the corresponding eigenvalue equation turns into
\begin{equation}\label{2.4}
\tilde{h}^{BR}\;\psi_g\;=\;\tilde{E}_g\;\psi_g.
\end{equation}
In order to prove Theorem \ref{t.1} we thus have to establish two items,
\begin{enumerate}
\item[(i)]
the existence of a ground state  of $\tilde{h}^{BR}$ for small $\tilde{m}\geq 0$ and  fixed $B_0>0$,
\item[(ii)]
the convergence of $\tilde{E}_g(\tilde{m})$  as $\tilde{m} \to 0$
to the ground-state energy $\tilde{E}_g(0)$ of $\tilde{h}^{BR}$ at $\tilde{m}=0.$
\end{enumerate}
As a consequence of (ii),
\begin{equation}\label{2.5}
\lim_{\mu_0 \to \infty} \,E_g/\mu_0^{\frac12}\;=\;\lim_{\tilde{m}\to 0} \tilde{E}_g(\tilde{m})\;=\;\tilde{E}_g(0),
\end{equation}
implying $E_g \sim \mu_0^{\frac12} \,\tilde{E}_g(0)\;=\;-c\,B^{\frac12}\,$ as $B \to \infty$
where $c:= -\tilde{E}_g(0)/B_0^{\frac12}.$

\section{Existence of the ground state}

\setcounter{equation}{0}

\begin{lemma} \label{l1}
Let $h^{BR}$ be the Brown-Ravenhall operator for an electron of mass $m$ in a constant magnetic field $B_0$. Then for $0.1\lesssim \gamma < \frac{2}{\pi}$ there exists $m_c >0$ such that $h^{BR}$
has a discrete ground state below $m$ for  $0\leq m<m_c$.
\end{lemma}

Equivalently, $h_-^{BR}:=h^{BR}-m$ has a discrete ground state below 0 for $m \in [0,m_c).$

\begin{proof}\quad\\
First we show that the spectrum of $h^{BR}$ is discrete below $m$, if nonempty.
From $E_A\geq m$ we have $\sigma(E_A) \subset [m,\infty)$
and hence $\sigma_{ess}(E_A) \subset [m,\infty).$
Actually $m \in \sigma_{ess}(E_A)$. This follows from $m \in \sigma_{ess}(D_A)\;$ (when $m>0$) for a constant magnetic field \cite[p.202]{Th}.
The spectrum is invariant if the unitary Foldy-Wouthuysen transformation $U_0$ is applied to  $D_A$,
which gives $U_0D_AU_0^{-1}=\beta E_A$ \cite{DeV} and proves the statement. Since the magnetic field does not restrict the electronic motion parallel to the field,
$\sigma(E_A)$ is continuous and hence $\sigma_{ess}(E_A)=[m,\infty).$ From (\ref{3.7a}) below it follows that this holds also for $m=0.$

For $m>0$ it was proven in \cite{Jaku06}  that $\sigma_{ess}(h^{BR})=\sigma_{ess}(E_A).$
The respective proof, using the compactness of the difference of the resolvents of $h^{BR}$ and $E_A$, has to be extended to the case $m=0$.
This is straightforward because the compactness property 
 relies basically on the invertibility of the operators $E_A+\mu$ and $h^{BR} +\mu$ for a suitable $\mu>0$
as well as on the relative form boundedness (\ref{1.4})
of the potential $V_1+V_2$ with respect to $E_A$ with form bound smaller than 1. For $m=0$, (\ref{1.4}) remains valid and thus the semiboundedness of $h^{BR}.$
An additional ingredient of the proof is the compactness of the operator $K:= \chi_0\,(E_A+\mu)^{-1}$ where $\chi_0$ is a bounded nonnegative 
function in coordinate space with $\chi_0 \to 0$ as $x \to \infty.$ Since an error occurred in \cite{Jaku06}, we give that proof anew for $m\geq 0.$

Let $S_A^2:= (\bfp-e\bfA)^2+m^2$ be the free Schr\"{o}dinger operator (increased by $m^2$) in a magnetic field and decompose
\begin{equation}\label{3.3}
K\;=\;[\chi_0\;(S_A+\mu)^{-1}]\;(S_A+\mu)\;(E_A+\mu)^{-1}.
\end{equation}
The compactness of the first factor can be shown with the help of the diamagnetic inequality. One has the pointwise inequality \cite{AHS1}
for any $\bfA \in L_{2,loc}({\Bbb R}^3)$ and $\psi \in L_2({\Bbb R}^3) \otimes {\Bbb C}^2$,
\begin{equation}\label{3.4}
|(\frac{1}{S_A+\mu}\;\psi)(\bfx)|\;\leq\; (\frac{1}{E_p+\mu}\;|\psi|)(\bfx)
\end{equation}
where $E_p=\sqrt{p^2+m^2}.\;$
Upon multiplication with $\chi_0$ it follows that
$|\chi_0(S_A+\mu)^{-1}\psi)(\bfx)|\,\leq\,(\chi_0(E_p+\mu)^{-1}\,|\psi|)(\bfx).\;$
The operator $\chi_0(E_p+\mu)^{-1}$ is compact as a product of two bounded functions of $x$ respective $p$ tending to zero at infinity. Therefore $\chi_0(S_A+\mu)^{-1}$ is compact, too \cite[Thm 2.2]{AHS1}.

Concerning the boundedness of the remaining factor in (\ref{3.3}) we have for $|\bfB|\,\leq B_0$ and $\psi \in H_{1/2}({\Bbb R}^3) \otimes {\Bbb C}^2,$
\begin{equation}\label{3.5}
\|S_A\;\psi\|^2\;=\;(\psi,(E_A^2+e\bfsigma\bfB)\;\psi)\;\leq\; \|E_A\,\psi\|^2\,+\,eB_0\;\|\psi\|^2,
\end{equation}
whence $\|S_A\psi\|\,\leq\,\|E_A\psi\|\,+\,\sqrt{eB_0}\,\|\psi\|.\;$
With $\psi:= (E_A+\mu)^{-1}\varphi$ it follows that
$$\|(S_A+\mu)\;\frac{1}{E_A+\mu}\,\varphi\|\;\leq\; \|(E_A\;\frac{1}{E_A+\mu})\,\varphi\|\,+\,\sqrt{eB_0}\;\|\frac{1}{E_A+\mu}\,\varphi\|\,+\,
\|\mu\,\frac{1}{E_A+\mu}\,\varphi\|$$
\begin{equation}\label{3.6}
\leq\; c\;\|\varphi\|
\end{equation}
with some constant $c$.

Having thus established that $\sigma_{ess}(h^{BR})=\sigma_{ess}(E_A)$ for $m\geq 0$ and $\gamma <\frac{2}{\pi}$, we have $\sigma_{ess}(h^{BR})=[m,\infty)$
under the same conditions.

\vspace{0.2cm}
Next we show that $\sigma(h^{BR}) \cap \,(-\infty,m)\,=\sigma_d(h^{BR})\neq \emptyset.$

\vspace{0.2cm}
\noindent{(a)$\;\;$ Case $m=0$}

In \cite{Jaku09} it was shown numerically for relativistic atoms ($Z\geq 20$; 
using the scaling and $\frac{eB_0}{2}=1)$ that there exists a trial function
$\psi_t \in L_2({\Bbb R}^3) \otimes {\Bbb C}^2$,
\begin{equation}\label{3.5a}
\psi_t(\bfx)\;=\;N_t\;e^{-eB_0\varrho^2/4}\;e^{-Z_{eff}\sqrt{x_3^2+1/(eB_0)}}\;{1 \choose 0}
\end{equation}
with $\varrho=\sqrt{x_1^2+x_2^2}$ and $N_t$ a normalization constant, and an effective charge $Z_{eff}>0$ 
 such that
\begin{equation}\label{3.7}
(\psi_t,h^{BR}\;\psi_t)\;<\;0 \qquad\quad \mbox{ for } m=0.
\end{equation}
Thus $\sigma_d(h^{BR})\neq \emptyset$ which
assures the existence of a ground state for a given nuclear charge $Z_0$.
A trial function for  $Z_0$  satisfying (\ref{3.7}) then also satisfies this inequality for all
$Z>Z_0.$
This follows from the fact that the negative part $-(V_1+V_2)$ of $h^{BR}\;\;
(V_1$ and $V_2$ are symmetric operators and hence are $\leq 0$ as is $V$)
 increases linearly with $\gamma$ (i.e. with  $Z$)
while the positive part $E_A$ is independent of $Z$.
We remark that a variational solution to (\ref{3.7}) can also be found  for $Z<20$, however the convergence of the respective integrals in $(\psi_t,h^{BR} \psi_t)$
gets increasingly poor when $Z$ becomes smaller.

\vspace{0.2cm}
\noindent{(b)$\;\;$ Case $m>0$}

We consider the operator $h^{BR}_-=h^{BR}-m$ 
where the subtraction of the rest energy $m$ has the advantage that $\sigma_{ess}(h^{BR}_-)=[0,\infty)$
implying that the ground-state energy is below zero for all $m$. We show that $(\psi_t,h^{BR}_-\,\psi_t)$ with the fixed $\psi_t$ from (\ref{3.5a}) is monotonically
decreasing with $m$ for $m <m_c$ when $m_c$ is sufficiently small.
For the kinetic part we have, using that $E_A\psi_t=\sqrt{p_3^2+m^2}\,\psi_t\;$ (see also section 4)
\begin{equation}\label{3.7a}
(\psi_t,(E_A-m)\;\psi_t)\;=\;(\hat{\phi}_t,(\sqrt{k^2+m^2}\,-m)\;\hat{\phi}_t),
\end{equation}
$$\hat{\phi}_t(k)\;=\;\tilde{N}\;\frac{K_1(\sqrt{Z_{eff}^2+k^2}/\sqrt{eB_0})}{\sqrt{Z_{eff}^2+k^2}},$$
where $\hat{\phi}_t \in L_2({\Bbb R})$ is (up to a constant) the Fourier transform of $\exp(-Z_{eff}\sqrt{x_3^2+1/eB_0})$, $\;K_1$ a modified Bessel function and $\tilde{N}>0$ a normalization constant \cite{Jaku09},\cite[p.482]{Grad}.
With the mean value theorem applied at $m=0$  we get
\begin{equation}\label{3.8}
(\hat{\phi}_t,(\sqrt{k^2+m^2}\,-m)\;\hat{\phi}_t)\;=\;(\hat{\phi}_t,|k|\,\hat{\phi}_t)\;-\;m\cdot D_t(\xi)
\end{equation}
with $\xi \in [0,m]$ and
\begin{equation}\label{3.9}
D_t(\xi):=\;2\tilde{N}^2\int_0^\infty dk\;\frac{K_1^2(\sqrt{Z_{eff}^2+k^2}\,/eB_0)}{Z_{eff}^2+k^2}\cdot \frac{k^2}{(\xi+\sqrt{k^2+\xi^2})\sqrt{k^2+\xi^2}}.
\end{equation}
Since the second factor in (\ref{3.9}) is nonnegative,  bounded by 1 and decreasing with $\xi$, the integrand is positive except for $k=0$ leading to $0<D_t(m_1)<D_t(\xi)\leq 1$ for all $\xi$ and $m$ below some fixed value $m_1$. Consequently,  (\ref{3.8}) is
 decreasing with $m$ (linearly for  $m \to 0)$.

Fourier transforming the expectation value of the potential part of $h^{BR}$, we have
\begin{equation}\label{3.10}
(\psi_t,(V_1+V_2)\;\psi_t)\;=\;-\gamma\, c_0\int_{-\infty}^\infty dk\int_{-\infty}^\infty dk' \;\hat{\phi}_t(k)\;\hat{V}_0(k-k')\;f_{kk'}(m)\;\hat{\phi}_t(k')
\end{equation}
$$f_{kk'}(m):=\;A_E(k)A_E(k')\,+\,A_E(k)\,\frac{k}{E_A(k)+m}\;\frac{k'}{E_A(k')+m}\;A_E(k')$$
where $c_0$ is a constant, $E_A(k)=\sqrt{k^2+m^2}$, $A_E(k)=(\frac{E_A(k)+m}{2E_A(k)})^\frac12$, and $\hat{V}_0(q)$ is the Fourier transformed expectation value of $\frac{1}{x}$ with the transverse function $e^{-eB_0\varrho^2/4}\;$ (for details see section 4, in particular (\ref{4.6})).

One can split $F(m):=(\psi_t,(V_1+V_2)\,\psi_t)$ into $F_+(m)+F_-(m)$ where $F_+(m)$ results from $kk'\geq 0$ in the
integrand of (\ref{3.10}) while $F_-(m)$ is assigned to $kk'\leq 0.$
The function $f_{kk'}$ is positive and  analytic for $m\geq 0$ if  $(k,k') \in {\Bbb R}^2 \backslash S_f$ with $S_f:=(0,{\Bbb R}) \cup({\Bbb R},0).\;$
Concerning $F_+(m),$ i.e. $kk'>0$, one gets by
 elementary computation  $\frac{df_{kk'}}{dm}|_{m=0}=0$  and $\frac{d^2f_{kk'}}{dm^2}|_{m=0} \neq 0\;$
(except for $k=k'$).
Thus $f_{kk'}$ has (almost everywhere) a local extremum in $m=0$.
Since  the remaining factors in the integrand of (\ref{3.10}) are independent of $m$ and positive
(see Lemma \ref{l2} for $\hat{V}_0),\;\,F_+(m)$ has also a local extremum
in $m=0$.

Thus there is an $m_c$ such that the variation of $F_+(m)$ in $[0,m_c)$ is considerably weaker than the linear dependence (\ref{3.8}) of the kinetic term.
The other contribution, $F_-(m)$, is monotonically decreasing like the kinetic term.
In fact, if $kk'<0,\;\;f_{kk'}$ is monotonically increasing with $m$
(as $1-\,\frac{|kk'|}{(E_A(k)+m)(E_A(k')+m)}$ is increasing with $m$).
As a result, $(\psi_t,(h^{BR}-m)\,\psi_t)$ is decreasing with $m$ in $[0,m_c).\;$
This feature is confirmed numerically, with  a quite large $m_c$.

Consequently, for a given nuclear charge $Z_0$, once $(\psi_t,h^{BR}_-\,\psi_t)<0$ is established at $m=0$ it follows that $(\psi_t,h^{BR}_-\,\psi_t)<0\;$ for $m \in [0,m_c).$ 
On the other hand, as discussed in the context of $m=0$ (but remains true for $m>0$),
$(\psi_t, h^{BR}_-\,\psi_t)<0$ for all charges greater than $Z_0$.
Since $Z_0\gtrsim10$ is arbitrary, this
guarantees the existence of
a ground state of $h^{BR}_-$ below $0$ for all $m<m_c.$
\end{proof}

Alternatively, the existence of a discrete ground state (and, in addition,  infinitely many bound states) of $h^{BR}$ below $m$ (for $m>0$)  may be based on a theorem by Matte and Stockmeyer \cite{MS}.
However, its applicability requires the fulfilment of certain nontrivial conditions on the
Weyl sequences for the essential spectrum of the free Dirac operator $D_A$.

\vspace{0.2cm}

In the remaining part of the present work we prove the convergence of the
sequence of eigenvalues $\tilde{E}_g(\tilde{m})$ for $\tilde{m} \to 0$ to $\tilde{E}_g(0).$
We start by restricting the ground-state function to the lowest Landau level and show continuity of the expectation value of $\tilde{h}^{BR}$ at $\tilde{m}=0$ for a certain class of functions.
Then we allow for the presence of higher Landau states.
From the continuity property of $\tilde{h}^{BR}$ the convergence of $\tilde{E}_g(\tilde{m})$ to $\tilde{E}_g(0)$ is deduced.

\section{Reduction to a one-dimensional problem}
\setcounter{equation}{0}

We want to gain information on the ground-state
wavefunction of $h^{BR}$  for arbitrary mass $m\geq 0.$
To this aim we reduce the three-dimensional problem to a one-dimensional one by invoking the eigenfunctions of the Pauli operator. The resulting eigenvalue equation
is then transformed to an integral equation in Fourier space from which the
basic properties of the ground-state function can be extracted.

The eigenfunctions $\psi_{nlds}(x_1,x_2)=\varphi_{nld}(x_1,x_2)\,\chi_s$ of the two-dimensional Pauli operator obey \cite[p.196]{Th}
$$(E_A^2-p_3^2-m^2)\;\psi_{nlds}\;=\;\left( \sum_{i=1}^2 (p_i-eA_i)^2\,-\,e\sigma_3B_0\right)\psi_{nlds}\;=\;(2n+1-s)\,eB_0\;\psi_{nlds}$$
\begin{equation}\label{4.1}
l\,\in\,{\Bbb N}_0,\qquad n \,\in \,{\Bbb N}_0,\qquad s\,=\,\pm 1,
\qquad d\;=\;\left\{ \begin{array}{rr}
1,& n=0\\
\pm 1,& n\geq 1
\end{array} \right.,
\end{equation}
where the spin functions are $\chi_{+1}={1 \choose 0},\;\chi_{-1}={0 \choose 1}.\;$
These eigenfunctions form a complete set of orthonormal functions (see e.g. \cite[\S111]{Lan},\cite{Ga}) such that an eigenstate $\psi \in L_2({\Bbb R}^3) \otimes {\Bbb C}^2$ of $h^{BR}$ can be expanded in terms of these functions.
When $B$ strongly dominates the Coulomb field the ground state of $h^{BR}$ is
 approximately characterized by the ground state of the Pauli operator which is determined by the quantum numbers $n=0$ and $s=+1$ (where the
rhs of (\ref{4.1}) attains its minimum zero).
Indeed it was shown for the Dirac operator in a homogeneous magnetic field \cite{DEL} that the $n=0$ approximation becomes exact when the nuclear charge tends to zero
or equivalently, when $B \to \infty$.

With the restriction to $n=0$ and $s=+1$ the (normalized) ground state $\psi$ of $h^{BR}$ can be written as
$$\psi_{n=0}(\bfx)\;=\;\sum_{l=0}^\infty   a_{l}\;N_l\;e^{-eB_0\varrho^2/4}\;\varrho^l\;e^{ild\varphi}\;\phi_{l}(x_3)\;\chi_{+1},$$
\begin{equation}\label{4.2}
N_l\;=\;\left( \frac{(eB_0)^{l+1}}{2^{l+1}l!\pi}\right)^\frac12,
\end{equation}
where $a_{l}$ is an  expansion coefficient, $\varrho=\sqrt{x_1^2+x_2^2},\;\;\phi_{l}(x_3)$ a normalized function yet to be
determined, and we have used the explicit form of $\varphi_{0l}$  \cite[p.196]{Th}, \cite[the negative sign of $d$ in that work relates to a positive coupling in (\ref{4.1}), $(p_i+eA_i)$]{RMS}.
When $n=0$ and the spin is fixed in $\psi$ one can show by using $\bfsigma \bfp_A \frac{1}{x} \bfsigma\bfp_A\,=\bfp_A \frac{1}{x}\bfp_A\,+i\bfsigma (\bfp_A\times \frac{1}{x}\bfp_A)$, that
not only $E_A$ and $V_1$, but also $V_2$ is diagonal in $l$.
For $s=+1$ we have from (\ref{4.1}) for the kinetic part
 $(\varphi_{0l},E_A\,\varphi_{0l})\,=\sqrt{p_3^2+m^2}=: E_A(p_3)$ and 
 it is straightforward to verify for the potential part
$$V_m(x_3,l):=\;(\chi_{+1}\,\varphi_{0l},-\frac{1}{\gamma}\,(V_1+V_2)\;\varphi_{0l}\,\chi_{+1})$$
\begin{equation}\label{4.3}
=\;A_E(p_3)\;\frac{(eB_0)^{l+1}}{2^l l!}\int_0^\infty d\varrho\;\varrho^{2l+1}\;e^{-eB_0\varrho^2/2}
\end{equation}
$$\cdot\left[ \frac{1}{\sqrt{\varrho^2+x_3^2}}\,+\,\frac{p_3}{E_A(p_3)+m}\;\frac{1}{\sqrt{\varrho^2+x_3^2}}\;\frac{p_3}{E_A(p_3)+m}\right] A_E(p_3)$$
where the functional dependence of $A_E(p_3)$ on $E_A(p_3)$ is given by (\ref{1.3}).
The smallest value of $(\psi, h^{BR}\psi)$ is given by $l=0$. In fact, $E_A(p_3)$ is independent of $l$. Moreover,
the weight factor of $(\varrho^2+x_3^2)^{-\frac12}$ in the integral in (\ref{4.3})
 is peaked at $\varrho_{max} = \sqrt{(2l+1)/eB_0}$, its normalized value at $\varrho_{max}$ slightly decreasing with $l$.
Thus $V_m(x_3,l)$ attains its maximum at $l=0$ where $\varrho_{max}$ is smallest.
For the consideration of the ground state we can therefore restrict ourselves to $l=0$.
Thus, dropping the sum in (\ref{4.2}) and the index on $\phi$,  we have $\psi_{n=0}(\bfx)=\varphi_{00}(x_1,x_2)\phi(x_3){ 1 \choose 0}$.
Disregarding the coupling to higher Landau states,
 the  eigenvalue equation for $h_-^{BR}$ reduces to
\begin{equation}\label{4.4}
\left[ \left( \sqrt{p_3^2+m^2}-m\right)\,-\,\gamma\;V_m(x_3,0)\right]\;\phi(x_3)\;=\;E_{m-}\;\phi(x_3).
\end{equation}
where $E_{m-}$ is the ground-state energy of $h^{BR} -m$ under the restriction $n=0$ (and $s=+1)$.

When Fourier transforming $\phi$ in (\ref{4.4}) and projecting with 
an eigenstate to $p_3,\;\,\exp(ipx_3)$,
the momentum operators in $V_m$ turn into functions of $k$ and $p$, respectively. As a result we obtain an
eigenvalue equation for the momentum-space function $\hat{\phi},$
$$\left( \sqrt{p^2+m^2}-m\right)\;\hat{\phi}(p)\,-\,\frac{\gamma}{\sqrt{2\pi}}\int_{-\infty}^\infty dk\;A_E(p)$$
\begin{equation}\label{4.5}
\cdot \left[ \hat{V}_0(p-k)\,
+\,\frac{p}{E_A(p)+m}\;\hat{V}_0(p-k)\;\frac{k}{E_A(k)+m}\right]
\;A_E(k)\;\hat{\phi}(k)\;=\;E_{m-}\;\hat{\phi}(p)
\end{equation}
for $ p \in {\Bbb R}$,
where we have introduced the Fourier transformed expectation value $V_0(x_3)$ of $\frac{1}{x}$
\cite[p.419]{Grad},
$$\hat{V}_0(q):=
\;\frac{eB_0}{\sqrt{2\pi}}\int_{-\infty}^\infty dx_3\;e^{-iqx_3}\int_0^\infty \varrho \,d\varrho\;e^{-eB_0\varrho^2/2}\;\frac{1}{\sqrt{\varrho^2+x_3^2}}$$
\begin{equation}\label{4.6}
=\;\frac{2eB_0}{\sqrt{2\pi}}\int_0^\infty \varrho\,d\varrho\;e^{-eB_0\varrho^2/2}\;K_0(|q|\varrho).
\end{equation}

\begin{lemma} \label{l2}
The momentum-space  potential $\hat{V}_0$ from (\ref{4.6}) obeys
\begin{equation}\label{4.7}
\hat{V}_0(q)\;\sim\left\{ \begin{array}{ll}
-c_1 \ln |q|,& q \to 0\\
&\\
\displaystyle\frac{c_2}{q^2},& |q| \to \infty
\end{array} \right.
\end{equation}
with $c_1>0$ and $c_2\geq 0.\;\;\hat{V}_0(q) \geq 0$ is monotonically decreasing with $|q|$.
\end{lemma}

\begin{proof}
For $q \to 0$ the modified Bessel function diverges logarithmically, $K_0(|q|\varrho) \sim - \ln (|q|\varrho) = - \ln |q|-\ln \varrho$,  and in this limit the integral over $\varrho$ is convergent.
For large $|q|$ we make the substitution $|q|\varrho =z$ and obtain
\begin{equation}\label{4.8}
\hat{V}_0(q)\;=\;\frac{2eB_0}{\sqrt{2\pi}\,q^2}\int_0^\infty z\,dz\;e^{-eB_0z^2/2q^2}\;K_0(z)\;\leq\; \frac{2eB_0}{\sqrt{2\pi}\,q^2},
\end{equation}
where we estimated the exponential by unity and used \cite[p.684]{Grad}. Since $K_0$ is monotonically decreasing and the integrand $\geq 0$, so is $\hat{V}_0(q).$
\end{proof}

With the help of Lemma 2 some properties of $\hat{\phi}(k)$ can be derived.
We have

\begin{lemma} \label{l3}
For $m\geq 0$ the restricted ground-state function in momentum space, $\hat{\phi}$, has the following properties
\begin{enumerate}
\item[(a)]
$\quad \hat{\phi}(k)$ is uniformly bounded for $k \in {\Bbb R}$.
\item[(b)]
$\quad |\hat{\phi}(k)|\;\leq \displaystyle\frac{c}{|k|},\quad |k| \to \infty,\;$ where $c\geq 0$ is some constant.
\end{enumerate}
\end{lemma}

\begin{proof}
For the proof of boundedness we use that $\sqrt{p^2+m^2}\,-m+\,|E_{m-}|\,\geq\,|E_{m-}|\;$ and estimate from (\ref{4.5}), using that $A_E\leq 1$ and $k/(E_A(k)+m)\,\leq 1$ (for $m\geq 0$),
$$|\hat{\phi}(p)|\;\leq\; \frac{2\gamma}{\sqrt{2\pi}}\;\frac{1}{|E_{m-}|}\int_{-\infty}^\infty dk\;\hat{V}_0(p-k)\;|\hat{\phi}(k)|$$
\begin{equation}\label{4.9}
\leq\; \frac{2\gamma}{\sqrt{2\pi}}\;\frac{1}{|E_{m-}|}\left( \int_{-\infty}^\infty dq\;\left( \hat{V}_0(q)\right)^2\right)^\frac12\;\|\hat{\phi}\|,
\end{equation}
where the Schwarz inequality was applied. Due to Lemma 2 the integral is bounded, and $\|\hat{\phi}\|=1,$ such that $|\hat{\phi}(p)|\,\leq \frac{c}{|E_{m-}|}$ where the constant $c$ does neither depend on $\hat{\phi}$ nor on $m$. Recall that $E_{m-}<0$ exists for all $m\geq 0$
since the trial function $\psi_t$ is an $n=0$ (spin-up)  state.

It is now easy to derive the behaviour of $\hat{\phi}(p)$ at $|p| \to \infty.$
Without restriction we can assume $|p|>m$ such that $\sqrt{p^2+m^2}-m+\,|E_{m-}|\,\geq \frac{p^2}{\sqrt{p^2+m^2}\,+m}\geq\,\frac{|p|}{\sqrt{2}+1}.$
For the estimate of the integral in (\ref{4.5}) we can use (\ref{4.9}). Then, with the same constant as above,
\begin{equation}\label{4.10}
|\hat{\phi}(p)|\;\leq\; \frac{\sqrt{2}+1}{|p|}\cdot c\qquad \quad \mbox{ for } |p|\,\to \infty.
\end{equation}
\end{proof}

\section{Continuity property of $h^{BR}$ in $m=0$}
\setcounter{equation}{0}

Let
\begin{equation}\label{5.1}
M_{pos}:=\;\{\varphi \in L_2({\Bbb R}):\;\|\varphi\|\,=1,\;\hat{\varphi} \mbox{ uniformly bounded near } 0\}
\end{equation}
be the set of square integrable functions for which there exists $C>0,\;\,0<\delta<1,$ such that $|\hat{\varphi}(k)|\leq C$ for $k \in [-\delta,\delta].\;$
Then for any $\epsilon_0>0$ there exists $0<\delta_0\leq \delta$ with
\begin{equation}\label{5.2}
2\;(\sup_{[-\delta,\delta]}|\hat{\varphi}|)^2\cdot 2\delta_0\;\leq\; 4C^2\,\delta_0\;\leq \;\epsilon_0.
\end{equation}
We can choose $\delta_0 = \min\{\frac{\epsilon_0}{4C^2},\delta\}$ for all $\varphi \in M_{pos}.$

\begin{lemma} \label{l4}
Let $h^{BR}(m)$ be the Brown-Ravenhall operator for a particle with mass $m$ and let $\psi(\bfx)=\,N_0\,e^{-eB_0\varrho^2/4}\phi(x_3)\,\chi_{+1}\,$ with $\phi \in M_{pos}.\;$
Then $(\psi, h^{BR}(m)\,\psi)$ is continuous in $m=0$, i.e. for any $\epsilon>0\;\,\exists\,m_0>0:$
\begin{equation}\label{5.3}
|(\psi,\left( h^{BR}(m)\,-\,h^{BR}(0)\right)\;\psi)|\;<\;\epsilon\qquad \mbox{ for all } m < m_0.
\end{equation}
\end{lemma}

As a consequence, the expectation value of $h_-^{BR}(m)$ is also continuous in 0.

\begin{proof}
Denoting the dependence on $m$ by an additional subscript we have for the kinetic part of $h^{BR}(m)$ from (\ref{1.3}),
\begin{equation}\label{5.4}
|(\psi,(E_{A,m}-E_{A,0})\;\psi)|\;\leq\; \|\psi\|\;\|\frac{m^2}{E_{A,m}+E_{A,0}}\;\psi\|\;\leq\;m\;\|\psi\|^2,
\end{equation}
which shows the continuity in $m=0.$
Note that (\ref{5.4}) holds for all $\psi \in L_2({\Bbb R}^3) \otimes {\Bbb C}^2.$

The potential part can be decomposed in the following way,
\begin{equation}\label{5.5}
V_{1,m}\,-\,V_{1,0}\;=\;-\gamma\,A_{E,m}\,\frac{1}{x}\,(A_{E,m}-A_{E,0})\;-\;\gamma\,(A_{E,m}-A_{E,0})\,\frac{1}{x}\,A_{E,0}
\end{equation}
with a similar formula for $V_2.\;$
The expectation value of the second summand is the conjugate complex one of the first factor (except for the replacement of $A_{E,m}$ by $A_{E,0})$ and need not be considered separately.
Transforming into Fourier space we have from (\ref{4.6})
$$|(\psi,(A_{E,m}-A_{E,0})\;\frac{1}{x}\;A_{E,m}\;\psi)|\;=\;|(\phi,(A_{E,m}(p_3)-A_{E,0}(p_3))\,V_0\,A_{E,m}(p_3)\;\phi)|$$
\begin{equation}\label{5.6}
\leq\; \int_{-\infty}^\infty dk\;|\hat{\phi}(k)|\int_{-\infty}^\infty dk'\;K(k,k')\;|\hat{\phi}(k')|
\end{equation}
with the kernel
\begin{equation}\label{5.7}
K(k,k'):=\;\frac{1}{\sqrt{2\pi}}\;|A_{E,m}(k)-A_{E,0}(k)|\;\hat{V}_0(k-k')\;A_{E,m}(k').
\end{equation}
We estimate (\ref{5.6}) further by applying the Schwarz inequality,
$$|(\psi,(A_{E,m}-A_{E,0})\frac{1}{x}A_{E,m}\psi)|\leq
\left( \int_{-\infty}^\infty dk|\hat{\phi}(k)|^2I(k)\right)^\frac12\left( \int_{-\infty}^\infty dk'|\hat{\phi}(k')|^2J(k')\right)^\frac12$$
\begin{equation}\label{5.8}
I(k):=\;\int_{-\infty}^\infty dk'\;K(k,k'),\qquad J(k'):=\;\int_{-\infty}^\infty dk\;K(k,k').
\end{equation}
We will show that each of the two factors can be bounded by an arbitrarily small $\epsilon^\frac12$,
provided $m$ is sufficiently small. We have, substituting $\xi=k-k'$ and using that $A_{E,m}(k')$ is bounded by 1,
\begin{equation}\label{5.9}
I(k)\;\leq\; \frac{1}{\sqrt{2\pi}}\;|A_{E,m}(k)-A_{E,0}(k)|\int_{-\infty}^\infty d\xi\;\hat{V}_0(\xi),
\end{equation}
where by Lemma \ref{l2} the integral is equal to some finite constant $c_I$.
We make use of the fact that $A_{E,m}(k)$ is continuous at $m=0$ if $k\neq 0$, choose $\epsilon_0>0$ and take $\delta,\;\delta_0$
from the definition (\ref{5.1}) of $M_{pos}$. Then for $|k|\geq \delta_0$ there is $m_{00} >0$ such that 
$|A_{E,m}(k)-A_{E,0}(k)|\,<\epsilon_0$ for all $m<m_{00}.$
We decompose the integration interval according to $(-\infty,\infty)=(-\infty,-\delta_0)\,\cup\,[-\delta_0,\delta_0]\,\cup\, (\delta_0,\infty)$ 
and obtain
$$\int_{-\infty}^\infty dk\;|\hat{\phi}(k)|^2\;I(k)\;\leq\; \frac{c_I}{\sqrt{2\pi}}\left\{\epsilon_0\left( \int_{-\infty}^{-\delta_0}+\int_{\delta_0}^\infty\right)dk\;|\hat{\phi}(k)|^2\right.$$
\begin{equation}\label{5.10}
+\;\left. \int_{-\delta_0}^{\delta_0}dk\;|\hat{\phi}(k)|^2\;|A_{E,m}(k)-A_{E,0}(k)|\right\}. 
\end{equation}
We have $|A_{E,m}(k)-A_{E,0}(k)|\,\leq 2,\;$ and since $\hat{\phi} \in M_{pos}$, by (\ref{5.2}) the last term in the curly brackets 
is estimated by $2\,( \sup\limits_{[-\delta_0,\delta_0]} |\hat{\phi}|\,)^2\cdot 2\delta_0\leq \epsilon_0.$
The other two terms can be estimated by $\epsilon_0 \,\|\hat{\phi}\|^2=\epsilon_0.$ This leads to
\begin{equation}\label{5.11}
\int_{-\infty}^\infty dk\;|\hat{\phi}(k)|^2\;I(k)\;\leq\; \frac{2c_I}{\sqrt{2 \pi}}\;\epsilon_0\qquad \mbox{ for } m<m_{00}.
\end{equation}

For  $J(k')$ we treat the case  $|k|\geq \delta_1$ with $\delta_1<1$ by estimating $|A_{E,m}(k)-A_{E,0}(k)|\,<\epsilon_0$ for, say,   $m<m_{01}$.
Then we obtain, substituting $\xi=k-k'$ for $k$,
$$J(k')\;\leq\; \frac{1}{\sqrt{2\pi}} \;A_{E,m}(k')\left\{ \epsilon_0\left( \int_{-\infty}^{-\delta_1-k'}+\int_{\delta_1-k'}^\infty\right)d\xi \;\hat{V}_0(\xi)\right.$$
\begin{equation}\label{5.12}
+\;\left. \int_{-\delta_1-k'}^{\delta_1-k'} d\xi\;|A_{E,m}(\xi+k')-A_{E,0}(\xi+k')|\;\hat{V}_0(\xi)\right\}.
\end{equation}
The first term in the curly brackets is estimated by $c_I\epsilon_0.\;$
As concerns the second term we profit from the fact that $\hat{V}_0(\xi)$ is symmetric and monotonically decreasing with $|\xi|$.
Thus, estimating the difference between $A_{E,m}$ and $A_{E,0}$ by 2, the remaining integral over $\hat{V}_0(\xi)$ has its maximum value for $k'=0$. With (\ref{4.7}) we have
$$\int_{-\delta_1-k'}^{\delta_1-k'}d\xi\;\hat{V}_0(\xi)\;\leq\; \int_{-\delta_1}^{\delta_1} d\xi\;\hat{V}_0(\xi)$$
\begin{equation}\label{5.13}
\leq\; -2c_1\int_0^{\delta_1} \ln \xi\;d\xi\;=\;2c_1\;\delta_1(1-\ln \delta_1).
\end{equation}
The rhs tends to zero as $\delta_1 \to 0$ and thus can be made smaller than $\epsilon_0$ for sufficiently small $\delta_1.$
Then, with $A_{E,m} \leq 1,$
\begin{equation}\label{5.14}
\int_{-\infty}^\infty dk'\;|\hat{\phi}(k')|^2\;J(k')\;\leq\; \frac{1}{\sqrt{2\pi}}\{ c_I\epsilon_0\,+\,2\epsilon_0\}\qquad \mbox{ for } m<m_{01}.
\end{equation}

The estimate of the second potential part, $V_{2,m} -V_{2,0}$ proceeds in the same way. One only has to replace throughout  $A_{E,m}(k)$  by $\tilde{A}_{E,m}(k):= A_{E,m}(k) \cdot \frac{k}{\sqrt{k^2+m^2}+m}$ which is also bounded by 1 and continuous at $m=0$ for $k \neq 0.$
This proves that the potential part is bounded by $\epsilon$ for $m<\min\{m_{00},m_{01}\}=:m_0.$
\end{proof}

We continue by showing that the state $\phi_m$, defining the ground state $\psi_{0,m}$ of both  $h^{BR}(m)$ and $h^{BR}_-(m)\;$
(under the restriction $n=0,\, s=+1$; the subscript $m$ is added for clarity), 
is a member of $M_{pos}$ for $m$ sufficiently small. For $m=0$ one has $\phi_0 \in M_{pos}$ because from (\ref{4.9}),
$|\hat{\phi}_0(p)|\,\leq \frac{c}{|E_0|}\;$ for all $p \in {\Bbb R}$ (where $E_0$ is the ground-state energy of the restricted $h^{BR}(0)$),
and we may choose any bound $C\geq \frac{c}{|E_0|}.\;$
Therefore, from Lemma \ref{l4} (with possibly a slightly smaller $m_0$),
\begin{equation}\label{6.1}
|(\psi_{0,0},h_-^{BR}(m)\;\psi_{0,0})-(\psi_{0,0},h^{BR}(0)\;\psi_{0,0})|\;<\;\epsilon\qquad \mbox{ for } m<m_0.
\end{equation}
As a consequence, $(\psi_{0,0},h^{BR}_-(m)\,\psi_{0,0})\,<\,E_0+\epsilon.\;$

On the other hand, $E_{m-}\leq\,(\psi_{0,0},h^{BR}_-(m)\,\psi_{0,0})\;$ since the expectation value of (the restricted) $h_-^{BR}(m)$ taken with an arbitrary function leads to an upper bound of the ground-state energy $E_{m-}.$
Combining these two inequalities we get
\begin{equation}\label{6.2}
E_{m-}\;<\;E_0\,+\,\epsilon\qquad\quad \mbox{ for } m<m_0.
\end{equation}
Since $E_0<0$ there exists $\delta_2<0$ such that $E_0<\delta_2<0.\;$
Let $\epsilon$ be so small that $E_0+\epsilon<\delta_2.\;$
Then, from (\ref{6.2}), $E_{m-}<\delta_2$ for all $m<m_0.$
This leads to an $m$-independent bound of $\hat{\phi}_m$ from (\ref{4.9}),
\begin{equation}\label{6.3}
|\hat{\phi}_m(p)|\;\leq\;\frac{c}{|E_{m-}|}\;<\;\frac{c}{|\delta_2|}\qquad \quad \mbox{ for } m<m_0.
\end{equation}
With the choice $C:= \frac{c}{|\delta_2|}\,>\,\frac{c}{|E_0|}\;$ we have found a universal bound on $\hat{\phi}_m$ for all $m<m_0.$
As a consequence, Lemma \ref{l4} holds for all eigenstates $\psi_{0,m}$ with $m<m_0.$

\vspace{0.2cm}

Let us now generalize the ansatz for the ground-state function $\psi$ of $h^{BR}$ by including the higher Landau states ($n>0$ and $s=-1$) in the expansion (\ref{4.2}).
This allows for the coupling of the potential $V_1+V_2$ to $n\geq 1$ states as well as for spin-flip induced by $V_2.$
However, for high magnetic fields the ground state is not much affected.
For the Dirac operator it was shown numerically \cite{DEL2} that for e.g. $Z=68$ and $B=2.3 \times 10^{16}$ G (where $Z^2\,m^2e^3\,=1.1 \times 10^{13}$ G is the particular field for which Coulombic and magnetic forces on the electron
are equally important) the effect of considering the $n>0$ contributions is about  3 percent.

With $\psi_{n=0}$ being  the ground state of $h^{BR}$ with the coupling 
to higher Landau states switched off, we now add a spin-flip term $\psi_{-1}$ and a remainder $\psi_r,$
\begin{equation}\label{5.15}
\psi(\bfx)\;=\;a_0\;\psi_{n=0}(\bfx)\;+\;a_1\;\psi_{-1}(\bfx)\;+\;a_2\;\psi_r(\bfx).
\end{equation}
$\psi_{-1}$ is characterized by $n=0,$ spin $s=-1$ and $l=1$ since $V_2$ changes $l$ by one unit simultaneously with the spin-flip.
$\psi_r$ is composed of Landau states with $n\geq 1$. All functions in (\ref{5.15}) are normalized and  mutually orthogonal.
The weight factors obey $|a_0|^2+|a_1|^2 +|a_2|^2=1$, guaranteeing that $\psi$ is normalized too.

Since $E_A$, acting on a Landau state with $n\geq 1$, is according to (\ref{4.1}) given by
$E_A=\sqrt{2neB_0 +eB_0(1-s)+p_3^2+m^2}\,\geq \sqrt{2eB_0+m^2}\;$ it is strictly positive at $m=0$.
The same is true when $E_A$ acts on an $n=0$ spin-down state,
resulting in $E_A=\sqrt{2eB_0+p_3^2+m^2}\,\geq \sqrt{2eB_0+m^2}.$
Hence, the $m$-dependent factors in $V_1$ and $V_2,\;\;A_E$ and $\bfsigma\bfp_A/(E_A+m),$
are analytic in $m=0$.
As a consequence, $(\psi_r,h^{BR}\psi_r)$ and $(\psi_{-1},h^{BR} \psi_{-1})$ are continuous in $m=0$ (in the sense
that $\psi_r$, respectively $\psi_{-1}$, is kept fixed when performing the limit $m \to 0$).

When proving Lemma \ref{l4} for $\psi$ from (\ref{5.15}) it thus remains to show
that the off-diagonal matrix elements are also bounded by $\epsilon$ for $m$ sufficiently small.
Considering the potential $V_1$ we have, using the decomposition (\ref{5.5}),
$$-\frac{1}{\gamma}\;(\psi_{n=0},(V_{1,m}-V_{1,0})\;\psi_r)\;=\;(\frac{1}{x^\frac12}\,A_{E,m}\,\psi_{n=0},\frac{1}{x^\frac12}\,(A_{E,m}-A_{E,0})\;\psi_r)$$
\begin{equation}\label{5.16}
+\;(\frac{1}{x^\frac12}\,(A_{E,m}-A_{E,0})\,\psi_{n=0},\frac{1}{x^\frac12}\,A_{E,0}\,\psi_r).
\end{equation}
The rhs of (\ref{5.16}) can be estimated from above by
$$\|\frac{1}{x^\frac12}\,A_{E,m}\,\psi_{n=0}\|\cdot \|\frac{1}{x^\frac12}\,(A_{E,m}-A_{E,0})\,\psi_r\|$$
\begin{equation}\label{5.17}
+\,(\psi_{n=0},(A_{E,m}-A_{E,0})\,\frac{1}{x}\,\tilde{A}_E\,\psi_{n=0})^\frac12\cdot (\psi_r, -\frac{1}{\gamma}\,V_{1,0}\,\psi_r)^\frac12
\end{equation}
with the bounded operator $\tilde{A}_E:=A_{E,m}-A_{E,0}.$
Since all Landau states, and thus $\psi_r$, have a Gaussian decay 
($\sim e^{-eB_0\varrho^2/4}$ \cite{Ga}), $(\psi_r,V_{1,0}\psi_r)$ is bounded.
The multiplication factor is bounded by $\epsilon^\frac12$ according to the proof of Lemma \ref{l4}.
The two factors in the first term of (\ref{5.17}) are bounded by $\epsilon$ since $(\psi_{n=0},V_{1,m}\,\psi_{n=0})$ is bounded and $A_{E,m}$ continuous in $m=0$ when acting on $\psi_r$.

The corresponding estimates for the potential $V_2$ are done in the same way.
All these estimates also hold when $\psi_r$ in (\ref{5.15}) is replaced by $\psi_{-1}$ or when $\psi_{n=0}$ is replaced by $\psi_{-1}.$
Together with (\ref{5.4})
this establishes the continuity property of $h^{BR}$, respectively $h_-^{BR}$,
\begin{equation}\label{5.22}
|(\psi,(h^{BR}_-(m)-h^{BR}(0))\,\psi)|\,<\epsilon\qquad \quad
\mbox{ for } m<m_0
\end{equation}
and $m_0$ sufficiently small, where $\psi$ from (\ref{5.15}) is taken as eigenstate to $h^{BR}(m')$ for any fixed $m'$ with $0\leq m'<m_0.$

\section{Convergence of the sequence of eigenvalues}
\setcounter{equation}{0}

With (\ref{5.22}) at hand it is easy to prove

\begin{lemma}\label{l5}
Let $E_{g-}(m)$ be the ground-state energy of $h_-^{BR}(m)=h^{BR}(m)-m,$ and let $E_g(0)$ be that of $h^{BR}(0).$
Then for a given $\epsilon>0$ there is an $m_0>0$ such that
\begin{equation}\label{6.4}
|E_{g-}(m)-E_g(0)|\;<\;\epsilon\qquad \mbox{ for all } m<m_0.
\end{equation}
\end{lemma}

\begin{proof}
Let us choose $\psi$ in (\ref{5.22}) as eigenstate $\psi_m$ to $h^{BR}(m)$. Then
\begin{equation}\label{6.5}
|(\psi_m,h_-^{BR}(m)\;\psi_m)\,-\,(\psi_m,h^{BR}(0)\;\psi_m)|\;<\;\epsilon\qquad \mbox{ for } m<m_0,
\end{equation}
such that $(\psi_m,h^{BR}(0)\,\psi_m)\,<E_{g-}(m)+\epsilon.\;$
Moreover, one has 
$E_g(0)\leq \,(\psi_m,h^{BR}(0)\,\psi_m)\;$ since $\psi_m$ will differ from the ground-state function $\psi_0$ of $h^{BR}(0).$
The combination of  these two inequalities leads to
\begin{equation}\label{6.6}
E_{g-}(m)\;>\;E_g(0)\,-\,\epsilon\qquad \mbox{ for } m<m_0.
\end{equation}
On the other hand, with $\psi:=\psi_0$, the step from (\ref{6.1}) to (\ref{6.2}) can be mimicked. Then
(\ref{6.2}) turns into $E_{g-}(m)<E_g(0)+\epsilon$ which gives,
combined with (\ref{6.6}),
\begin{equation}\label{6.7}
|E_{g-}(m)\,-\,E_g(0)|\;<\;\epsilon\qquad \mbox{ for } m<m_0.
\end{equation}
This proves the continuity of $E_{g-}(m)$ (and of $E_g(m)$ as well) at $m=0.$
\end{proof}

\section*{Acknowledgment}

It is a pleasure to thank P.M\"{u}ller for enlightening discussions.

\end{document}